\documentstyle[sprocl]{article}

\input{psfig}

\def\Journal#1#2#3#4{{#1} {\bf #2}, #3 (#4)}

\def\NPB{{\em Nucl. Phys.} B}
\def\PLB{{\em Phys. Lett.}  B}

\def\PRD{{\em Phys. Rev.} D}

\def\be{\begin{equation}}
\def\ee{\end{equation}}
\def\bea{\begin{eqnarray}}
\def\eea{\end{eqnarray}}
\def\sd{\hat s}
\def\tcd{\cos \hat \theta}
\def\lt{\lambda_t}
\def\ltb{\lambda_{\bar t}}

\begin{document}

\title{LOOP-INDUCED SUSY EFFECTS IN STRONG TOP PAIR PRODUCTION
\footnote{Talk presented at PASCOS-98, Boston, MA, March 1998.}}

\author{D. WACKEROTH}

\address{Theory Group F1, Paul Scherrer Institut,\\
CH-5232 Villigen PSI, Switzerland\\E-mail: wackeroth@psi.ch} 

\maketitle\abstracts{We present some results of 
the calculation of MSSM one-loop contributions to
strong $t \bar t$ production in hadronic collisions.
Here we concentrate on the MSSM
electroweak-like ${\cal O}(\alpha)$ and SQCD ${\cal O}(\alpha_s)$ 
corrections to the main production mechanism at the
Fermilab Tevatron $p \bar p$ collider, $q\bar q\to g \to t \bar t$.
We give results for the total $t \bar t$ production rate and
for parity violating
asymmetries in the production of left and right handed top quark pairs.}
\section{Introduction}
The potential of future hadron colliders 
to perform precision measurements of top quark observables~\cite{topacc} 
is the motivation for an ongoing study on loop effects in strong
$t \bar t$ production within the Electroweak 
Standard Model (SM)~\cite{diplpubsm,smasym} and 
beyond~\cite{topmssm,mssmasym,sqcd,asym}. 
While the SM electroweak (EW) corrections have only marginal effects on
$t \bar t$ observables such as the total $t \bar t$ production rate 
$\sigma(S)$, 
the invariant $t \bar t$ mass distribution $d\sigma/d M_{t \bar t}$
and parity violating asymmetries ${\cal A}_{LR}$ in the production of
left and right handed top quark pairs,
there is the possibility of considerable enhancements
within supersymmetric models. 
The ongoing study of radiative corrections to these observables within the
Minimal Supersymmetric extension of the SM (MSSM) 
reveals promisingly large effects at both the upgraded Tevatron and the LHC. 
At the LHC, for instance, the MSSM EW-like corrections typically
diminish the leading order production cross section $\sigma_B$ by 
$\stackrel{<}{\sim}10\%$, 
they can significantly distort $d\sigma/dM_{t \bar t}$
and they induce asymmetries of ${\cal A}_{LR} \stackrel{<}{\sim} 3\%$. 
In this contribution we will concentrate on 
combining the MSSM EW-like ${\cal O}(\alpha)$ and SQCD ${\cal O}(\alpha_s)$ 
corrections to the main production mechanism at the
Fermilab Tevatron $p \bar p$ collider, $q\bar q\to g \to t \bar t$.
We will present first results 
for $\sigma(\sqrt{S}=2$ TeV) and ${\cal A}_{LR}$. 
\section{Loop-induced SUSY effects in $p \bar p,pp \to t\bar t X$}
At the parton level, the next-to-leading order differential cross sections 
for polarized $t \bar t$ production via 
the $q \bar q$ annihilation and gluon fusion subprocesses are obtained
by contracting the corresponding matrix elements describing the
loop contributions
$\delta {\cal M}_i (i=q\bar q,gg)$~\cite{diplpubsm,topmssm} with the 
Born matrix elements ${\cal M}_B^i$
\be
\frac{d \hat \sigma_i(\hat t,\hat s,\lt,\ltb)}{d \tcd}
= \frac{\beta_t}{32 \pi \hat s} \, \overline{\sum}
\left[ \mid {\cal M}^i_B \mid^2+
2 {\cal R}e \; \overline{\sum}
(\delta {\cal M}_i \times {\cal M}_B^{i*})\right]+\mbox{higher order},
\label{eq:dsig}
\ee
where $\beta_t$ is the top quark velocity and $\lambda_{t,\bar t}$ denote the
$t (\bar t)$ helicity states. $\sd, \hat t$ are Mandelstam variables 
and $\hat \theta$ is the scattering angle of the top quark in the parton CMS.
$\delta {\cal M}_{q \bar q}$ comprises the  
EW(-like) ${\cal O}(\alpha)$ and SQCD ${\cal O}(\alpha_s)$ corrections.
So far we only included EW(-like) corrections to the gluon fusion subprocess
described by $\delta {\cal M}_{gg}$;
the SQCD calculation is work in progress~\cite{sqcd}.
We studied the effects of EW(-like) corrections 
on unpolarized $t \bar t$ observables in~\cite{diplpubsm,smasym,topmssm}.
Parity violating effects in the production of left and right 
handed top quark pairs
which are loop induced through EW(-like) interactions 
are discussed in~\cite{smasym,mssmasym,asym}.
In the following calculation of the SQCD ${\cal O}(\alpha_s)$ corrections 
we will closely follow the notation of~\cite{topmssm,sqcd} 
and refer to it for more details.
\subsection{SQCD ${\cal O}(\alpha_s)$ corrections to 
$q \bar q \to g \to t \bar t$}\label{subsec:sqcd}
\begin{figure}[t]
\hfill
\psfig{figure=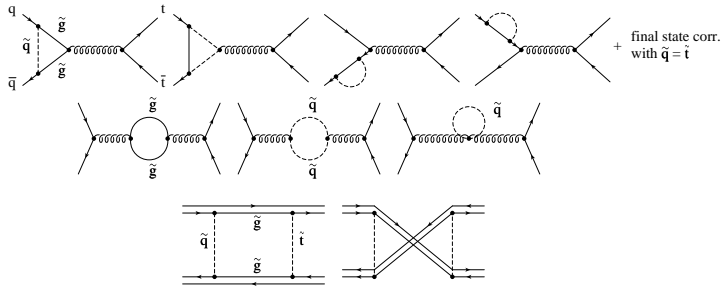,height=1.5in}
\caption{The Feynman diagrams to the SQCD ${\cal O}(\alpha_s)$ contribution to 
$q \bar q \to g \to t \bar t$.}
\label{fig:feyn}
\end{figure}

\begin{figure}
\hfill
\psfig{figure=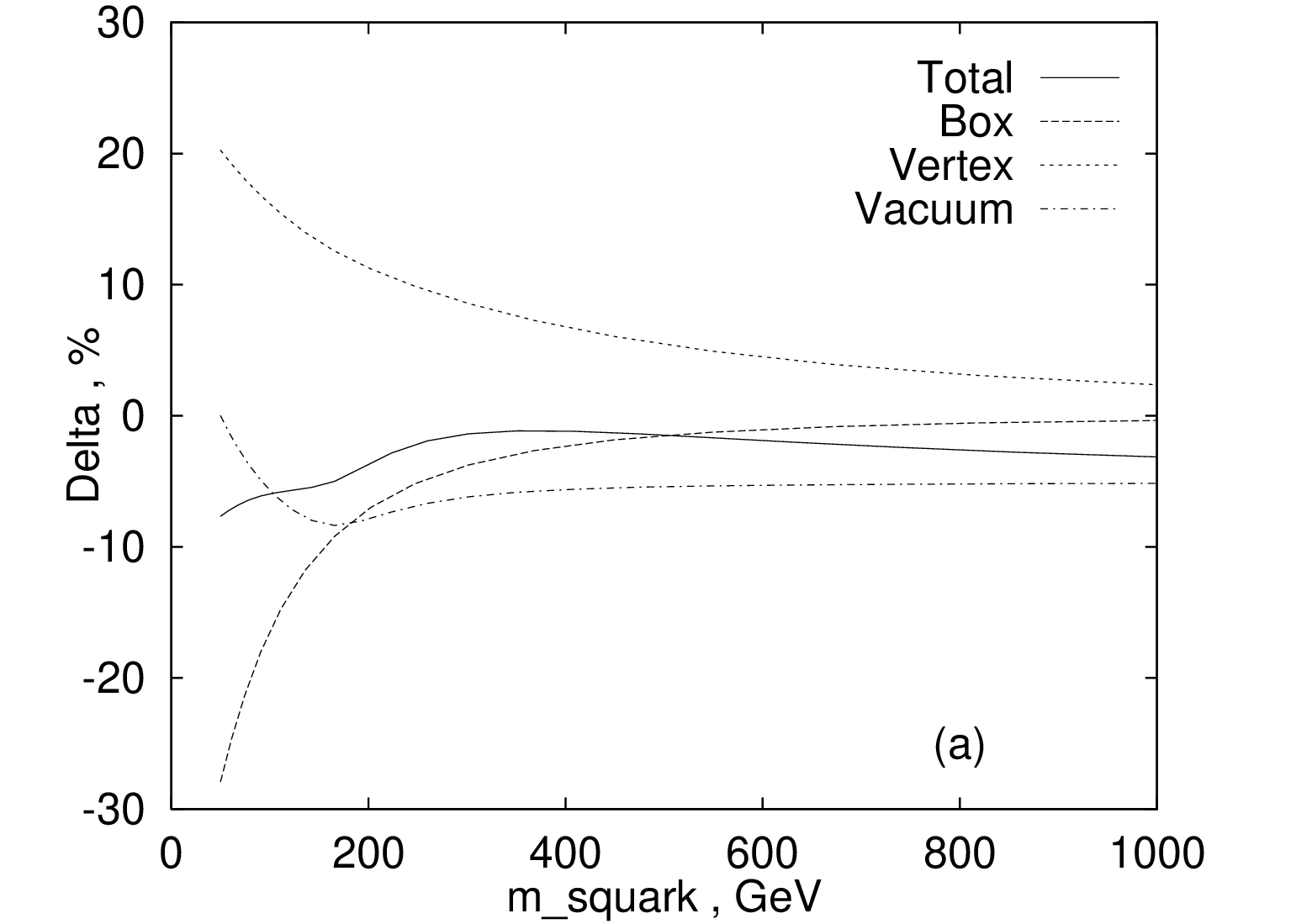,height=1.5in} \hfill
\psfig{figure=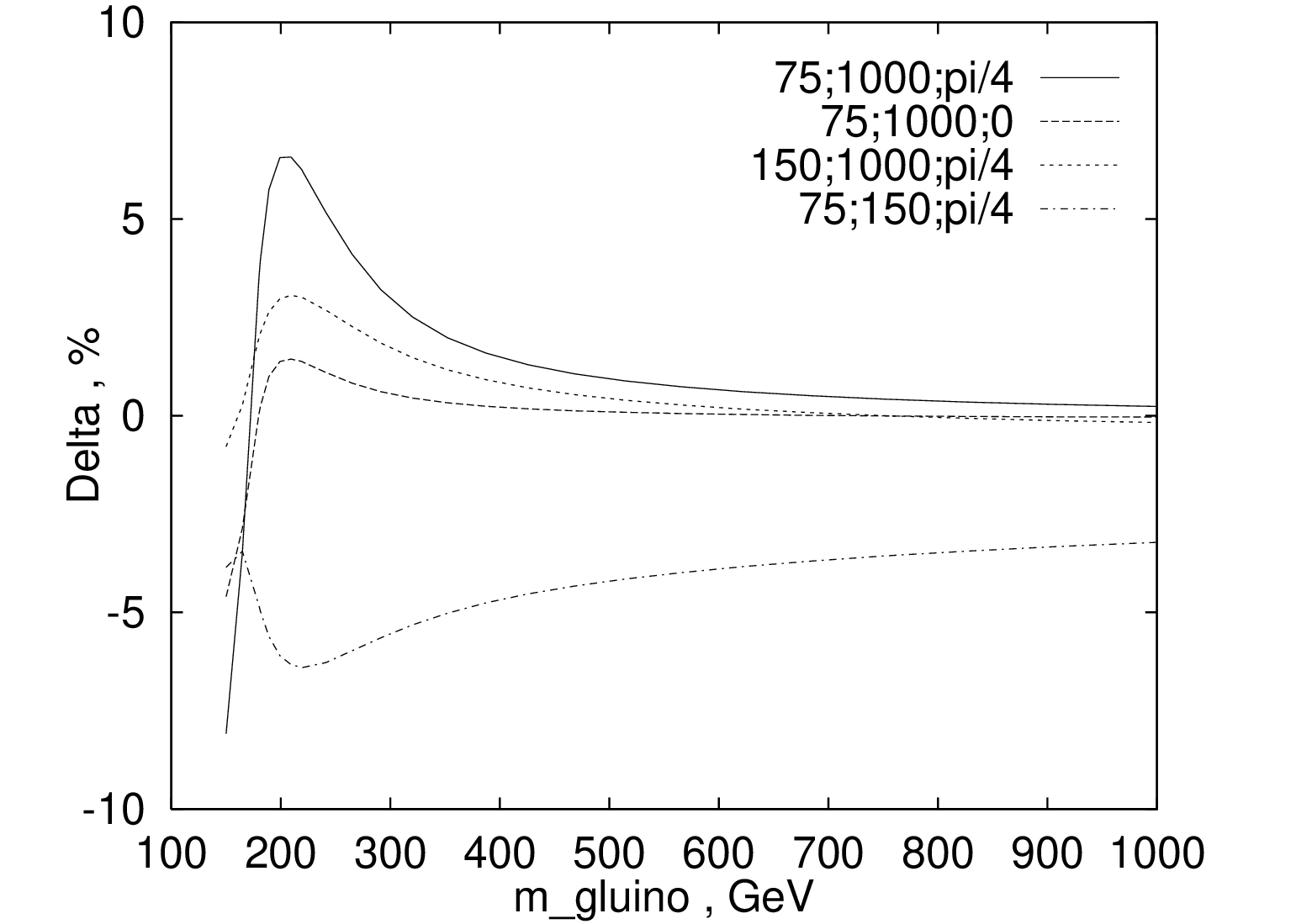,height=1.5in}
\caption{(a) The different contributions to the
SQCD ${\cal O}(\alpha_s)$ correction in dependence of a common squark mass
with $m_{\tilde g}=200$ GeV. (b) $\Delta$
in dependence of $m_{\tilde g}$ for different choices of $m_{\tilde t_2}$,
$m_{\tilde b_L}$ and $\Phi_{\tilde t}$ (with $m_t=174$ GeV).}
\label{fig:sqcd}
\end{figure}

In Fig.~\ref{fig:feyn} we display the Feynman diagrams contributing 
to the SQCD ${\cal O}(\alpha_s)$ corrections to $q \bar q \to g \to t \bar t$.
The corresponding contribution to $d \hat \sigma_{q\bar q}$ 
of Eq.~\ref{eq:dsig} reads ($z=\tcd$)~\cite{sqcd}
\bea
\lefteqn{ 2 {\cal R}e \; \overline{\sum}
(\delta {\cal M}_{q\bar q}^{SQCD} \times {\cal M}_B^{q\bar q*}) = }
\nonumber\\
& & \overline{\sum} \mid {\cal M}^{q\bar q}_B \mid^2
\frac{\alpha_s}{2 \pi} {\cal R}e
\left(F_V(\sd,m_q)+F_V(\sd,m_t)+\hat \Pi(\sd)\right)+
\nonumber\\
& & \frac{4 \pi \alpha_s^3}{9} {\cal R}e 
\left(\beta_t^2 (1-z^2) (1+4 \lt \ltb)
 F_M(\sd,m_t) + 2 (\lt-\ltb) \left[2 z G_A(\sd,m_q)+
\right. \right. \nonumber\\
& & \left. \left. \beta_t (1+z^2) G_A(\sd,m_t) \right]\right)
+ \frac{32 \pi \alpha_s^3}{9 \sd} \frac{1}{4} {\cal R}e 
\left( (-1) \frac{7}{3} B_t-\frac{2}{3} B_u \right)(\sd,\hat t) \; ,
\eea
where the vertex corrections 
are parametrized in terms of UV finite form factors $F_V,F_M$ and $G_A$,
$\hat \Pi(\sd)=\Pi(\sd)-\Pi(0)$ denotes the 
subtracted gluon vacuum polarization and
$B_t$ and $B_u$, respectively, parametrize the $t$ and $u$ channel 
box contributions. Their explicit expressions will be provided in~\cite{sqcd}.
In Fig.~\ref{fig:sqcd}(a) we show separately their numerical impact on
the total $t\bar t$ production rate. We observe large cancellations between
the vertex and box contributions
\footnote{Since we differ from~\cite{zack}
in the overall sign of the box diagrams
we feel a brief explanation is in order:~to determine 
the relative sign between the $t$ and $u$ channel box diagram
(as well as between box and Born contribution)
we apply the rules of~\cite{dehk}.
Fixing the reference order as $t \bar t \bar q q$ and choosing the 
fermion flow as indicated in Fig.~\ref{fig:feyn} we only
need to assign an additional minus sign to the $t$ channel box contribution.}.
\begin{figure}
\hfill
\psfig{figure=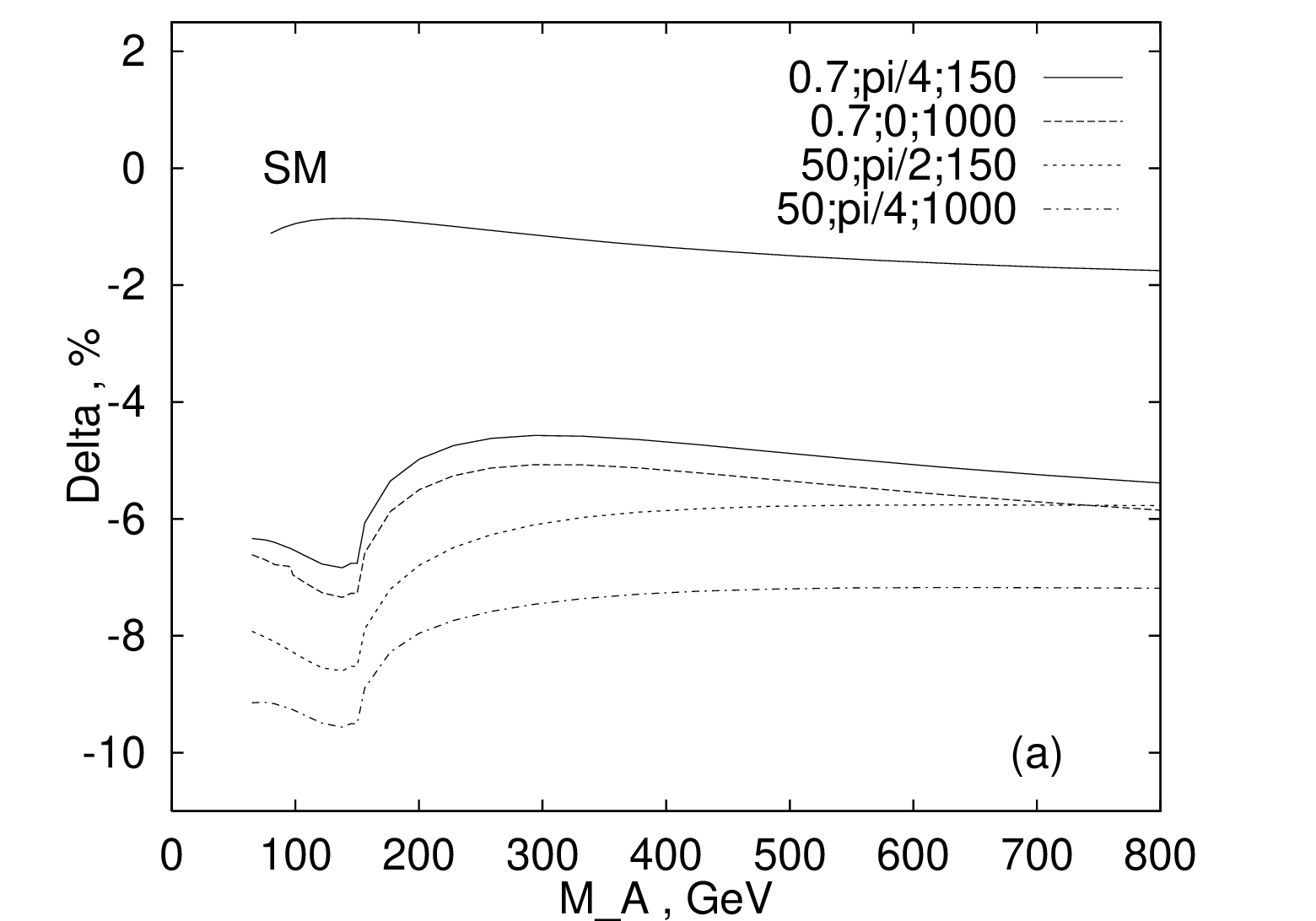,height=1.5in} \hfill
\psfig{figure=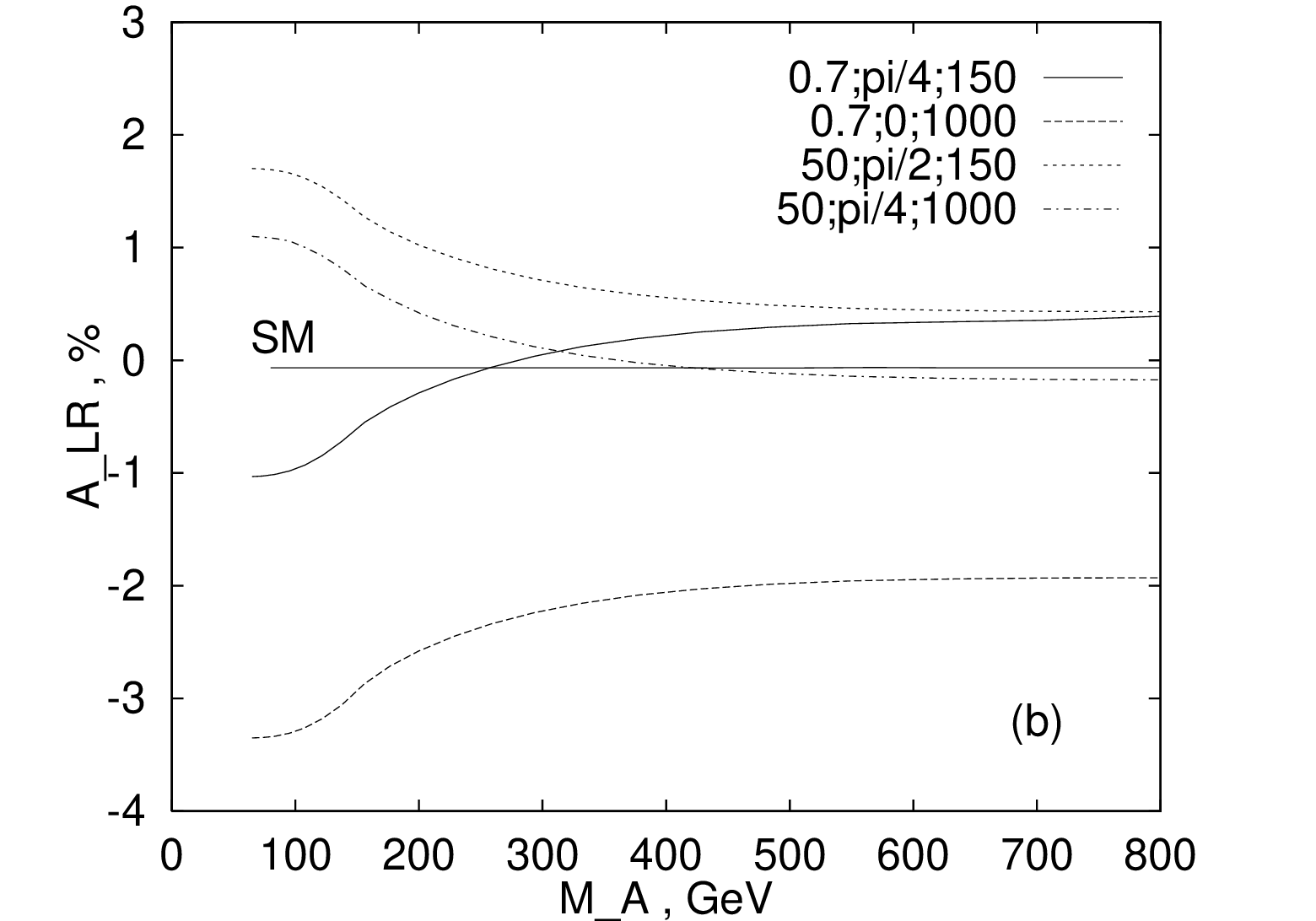,height=1.5in}
\caption{(a) $\Delta$ and (b) ${\cal A}_{LR}$ including the
combined EW-like ${\cal O}(\alpha)$ and 
SQCD ${\cal O}(\alpha_s)$ corrections in dependence of $M_{A^0}$
for different choices of $\tan\beta$, $\Phi_{\tilde t}$ and
$m_{\tilde b_L}$ (with $m_t=174$ GeV and $m_{\tilde t_2}=90$ GeV, 
$\mu=-120$ GeV, $M_2=3 |\mu|$, $m_{\tilde g}=150$ GeV).} 
\label{fig:tot}
\end{figure}
\subsection{Loop-induced EW-like and SQCD effects 
in $p \bar p\rightarrow t \bar t X$ at the Tevatron}\label{subsec:teva}
The hadronic cross section $\sigma_{\lt,\ltb}(S)$
to polarized $t \bar t$ production is obtained by
convoluting $\hat \sigma_i$ of Eq.~\ref{eq:dsig}
with the corresponding parton distribution functions.
We introduce a relative correction $\Delta=\sigma/\sigma_B-1$
(with $\sigma=\sum_{\lt,\ltb} \sigma_{\lt,\ltb}$)
which reveals the numerical impact of the radiative corrections on
the total $t \bar t$ production rate. In Fig.~\ref{fig:sqcd}(b)
we show $\Delta$ in dependence of
the gluino mass $m_{\tilde g}$ for different choices 
of the light stop quark mass $m_{\tilde t_2}$, the sbottom 
quark mass $m_{\tilde b_L}$ and the stop mixing angle $\Phi_{\tilde t}$. 
While the MSSM EW-like corrections are only of the order
of a few $\%$ (apart from the threshold region 
$m_t \approx m_{\tilde t_2}+M_{\tilde\chi^0}$ 
where they can reach $30 \%$)~\cite{topmssm} the
SQCD one-loop corrections can considerably diminish/enhance 
$\sigma_B$ when the gluino is not too heavy.
We now combine the MSSM EW-like corrections of~\cite{topmssm} with 
the SQCD corrections of Sec.~\ref{subsec:sqcd}. 
The resulting relative correction $\Delta$ 
and integrated left-right asymmetry 
${\cal A}_{LR}=[\sigma_{\lt=+1/2,\ltb=-1/2}
-\sigma_{-+}]/[\sigma_{+-}+\sigma_{-+}]$, respectively,
are shown in Fig.~\ref{fig:tot}(a) and (b) for different choices of the
MSSM input parameters. For comparison we also display the SM results
when varying the SM Higgs boson mass.
The SQCD ${\cal O}(\alpha_s)$ contribution induces parity violating asymmetries
when the squarks are non-degenerate in mass (and
$\Phi_{\tilde t} \ne \pi/4$). Then the asymmetries
induced by EW(-like) corrections can be considerably enhanced and are
possibly observable\footnote{
When summing over the $\bar t$ helicities the resulting asymmetry
${\cal A}$ is statistical significant 
($N_s \ge 4$~\cite{smasym,mssmasym}) when  
$|{\cal A}| \stackrel{>}{\sim}1.8\%$ 
which corresponds to $|{\cal A}_{LR}| \stackrel{>}{\sim}2.4\%$.}
at the upgraded Tevatron with ${\cal L}=10\, \mbox{fb}^{-1}$.

\section*{Acknowledgments}
This work is done in collaboration with W. Hollik, C. Kao and W.M. M{\"o}sle.

\end{document}